\input{psfig.sty}
\documentclass[11pt,a4paper]{article}
\usepackage{jcappub}

\title{Cusps and Cores in the presence of galactic bulges}
\author[a,b]{A. Del Popolo}
\author[c]{N. Hiotelis}
\affiliation[a]{Dipartimento di Fisica e Astronomia, University Of Catania, \\
Viale Andrea Doria 6, 95125 Catania, Italy \\}
\affiliation[b]{International Institute of Physics, Universidade Federal do Rio Grande do Norte,\\
59012-970 Natal, Brazil}
\affiliation[c]{1st Experimental Lyceum of Athens, Ipitou 15, Plaka, GR-10557 Athens, Greece}

\emailAdd{adelpopolo@oact.inaf.it}
\abstract{In this paper, we study how the presence of bulge formation in galaxies influence their inner density profile, by means of an extended version of the Del Popolo (2009) semi-analytical model. As in Del Popolo (2009), the model takes into account the effect of baryons adiabatic contraction, ordered and random angular momentum, dynamical friction, and adds to the previous the effect of gas cooling, star formation, supernova feedback, and reionization. Our model shows that dwarf galaxies are bulgeless, in agreement with observations showing that the large majority of them has no stellar bulges, and are characterized by a flat profile well described by a Burkert profile. 
We then studied the effect of a bulge, added to the cored DM halo, on the density profile. In the case of a galaxy having a mass $10^{11} M_{\odot}$ the inner density profile has a slope $\alpha \simeq 0.65$, for a bulge of $4.5 \times 10^{9} M_{\odot}$ , while if bulge formation is not considered, the slope would be $\alpha \simeq 0.55$. If the bulge is larger, $6.5 \times 10^{9} M_{\odot}$ the slope is $\alpha \simeq 0.7$. In the case of a larger galaxy with $10^{12} M_{\odot}$ the 
slope is $\alpha \simeq 0.85$, while in absence of bulge it is $\alpha \simeq 0.75$. 
We finally study how the inner slope $\alpha$ changes with the bulge mass, and we find a correlation among the two quantities. The result shows that bulge formation has an important role in shaping the inner DM density profile in agreement with Inoue \& Saitoh (2011) result. The result implies that going from Sc to SO Hubble type the slope is slightly steepening due to the bulge formation, and due to the fact that early type galaxies have larger bulges. }
\begin{document}
\maketitle

\section{Introduction}

N-body simulations of density profiles forming in the $\Lambda$CDM model result in cuspy profile in haloes centers 
(e.g. Navarro, Frenk, \& White 1996, 1997 (NFW); Fukushige \& Makino 2001; Klypin et al. 2001; Navarro et al. 2010; Springel et al. 2008; Navarro et al. 2010).
However, it is long known that the profile of dwarfs and LSB galaxies have a cored halo (Burkert 1995; de Blok, Bosma, \& McGauch 2003; Del Popolo 2009 (DP09) , Del Popolo \& Kroupa 2009; Del Popolo 2012a,b (DP12a, DP12b); Del Popolo \& Gambera 2000; Oh et al. 2010, 2011; Kuzio de Naray \& Kaufmann 2011). Cosmological\footnote{Examples are: a) perturbations power spectrum modification (Zentner \& Bullock 2003); b) modification of the particles constituting DM (Colin, Avila-Reese \& Valenzuela 2000;  Hu, Barkana \& Gruzinov 2000; Goodman 2000; Peebles 2000; Kaplinghat, Knox, \& Turner, 2000); c) modified gravity theories (Buchdal 1970; Milgrom 1983; Ferraro 2012).}, and astrophysical solutions\footnote{Examples are: a) cusp into core transformation trough supernovae feedback (Navarro, Eke, \& Frenk 1996; Read \& Gilmore 2005; Maschchenko, Wadsley \& Couchman 2008; Governato et al. 2010; Governato et al. 2012), and b) transfer of angular momentum from baryons to DM (El-Zant et al. 2001, 2004; Romano-Diaz et al. 2008, 2009; Del Popolo 2009; Cole et al. 2011).} have been proposed to solve the quoted problem.

Several of the previous studies are concentrated on dwarf galaxy scales, except some studies (e.g., El-Zant et al. 2004; DP09; Governato et al. 2012)  
%(except El-Zant et al. 2004, DP09, DP12a, Governato et al. 2012)  
and this is because dynamical perturbations affects more the quoted systems having a shallow potential in comparison with more massive system (DP12b). 

While Dwarfs and LSBs are characterized by high values of the $M/L$ ratio even in the central parts, implying that DM dominates on baryons, 
%which (baryons) can be used to trace the DM distribution, 
many galaxies are dominated by baryons in their centers. This, as expected, produces a noteworthy change in DM distribution. In the case of high-surface brightness (larger mass objects) the determination of the structure of the inner density profile is more complicated than dwarf galaxies. While Span\'o et al. (2008) concluded that disc galaxies with high-surface brightness, from Sab to IM Hubble types, are characterized by a cored profile, different conclusions were reached by other authors (e.g., de Blok et al. 2008, Del Popolo \& Cardone 2012; Del Popolo, Cardone \& Belvedere 2013). By using the THINGS sample, de Blok et al. (2008) concluded that the low luminosity galaxies with $M_B>-19$ have profiles that are better described using the ISO model, while galaxies with $M_B<-19$ 
equally well described by cored or cuspy profiles. Even in the case of dwarfs, Simon et al. (2005) showed that the sample of galaxies NGC 2976, 4605, 5949, 5693, 6689, have inner slopes in the range 
$0$ (NGC 2976)$, -1.28$ (NGC 5963).

High redshift galaxies are characterized by clumpy stellar structures, with masses $10^{5}-10^{8} M_{\odot}$, (van den Bergh et al. 2006;
%Elmegreen, Elmegreen \& Hirst 2004; 
Elmegreen et al. 2009; Genzel et al. 2011), which are connected to the existence of a gas rich disc, and to the instability in gas accretion (Bournard, Elmegreen, and Elmegreen 2007; Agertz, Teyssier \& Moore 2009; Ceverino et al. 2011). According to the previous studies, current disc galaxies should be the result of the evolution of the previous structures. As previous discussed, El-Zant et al. (2001, 2004), DP09, Romano-Diaz et al. (2008, 2009), Cole et al. (2011), showed that clumps infall into the center of the galaxy will produce a flattening of the profile from a cuspy NFW profile, characterized by an inner density $\rho \propto r^{-1}$, to a Burkert profile, characterized by a central core.  

{Recently Inoue \& Saitoh (2011) (IS11), claimed that the infall of the clumps produces a core in the inner density profile,
%a flattening of the inner density profile, 
but if in the galaxy a bulge is formed, the cusp is re-formed. 
%This is in disagreement with several papers in the literature (e.g., El-Zant et al. 2001, 2004; El-Zant \& Shlosman 2002; Goerdt et al. 2010; Cole 2011) %in which the cusp does not revive, after the core is formed. In particular, Fu et al. (2007) found  that bulge is very slightly 
%influencing the DM halo.

IS11 discussed the cusp reformation after core creation in the context of clumpy disc formation followed by bulge formation (Noguchi 1998; Inoue \& Saitoh 2012). 
%(e.g., El-Zant et al. 2001, 2004; El-Zant \& Shlosman 2002; Goerdt et al. 2010; Cole 2011)
At high redshift, some galaxies, dubbed ``clump clusters" and ``chain galaxies" (e.g. van den Bergh et al. 1996;
Elmegreen et al. 2004, 2009; Genzel et al. 2011), show clumpy stellar structure. According to numerical simulations, the clumpy structure form due to the instability of the accreting gas (e.g. Noguchi 1998, 1999; Bournaud et al. 2007; Agertz et al. 2009; Ceverino et al.
2010, 2011), and then the clumpy galaxies would evolve into disc galaxies. Because of dynamical friction of baryons and DM, the clumps collapse in the galactic center, so heating the cuspy DM halo and forming a core, which is a transient state. When the bulge forms, the cuspy profile is revived.

In the present paper, we want to study by means of DP09 semy-analytical model the previous issue, namely how the bulge formation influence the density profile.}

In Sect. 2, we summarize the model, in Sect. 3 we present the results and discussion, and finally in Sect. 4, the discussion and summary.

\section{Model}

In this section, we shortly summarize the model used, already introduced  
%previously described 
in DP09, DP12a,b, and additionally describe the way reionization, cooling, star formation, and the supernova feedback are taken into account. 

The quoted model is an improvement of previous spherical (secondary) infall models (SIMs) (e.g., Gunn \& Gott 1972; Hoffman \& Shaham 1985; Le Delliou, \& Henriksen 2003; Ascasibar, Yepes \& G\"ottleber 2004; Williams, Babul \& Dalcanton 2004; Cardone, Leubner \& Del Popolo 2011). Differently from previous papers on SIMs, which considered, one at a time, just the radial collapse (Gunn \& Gott 1972), or the effects of non-radial collapse produced by random angular momentum (e.g., Ryden \& Gunn 1987)\footnote{The random angular momentum arise from the random velocities present in the system (Ryden \& Gunn 1987).}, or the adiabatic contraction of DM produced by baryon infall (Blumenthal et al. 1986; Gnedin et al. 2004; Gustafsson et al. 2006), dynamical friction (e.g., El-Zant et al. 2001, 2004), in 
DP09, DP12a,b, random and ordered angular momentum, dynamical friction, adiabatic contraction, 
%gas cooling, and star formation 
were all simultaneously taken into account\footnote{In Del Popolo, Pace, \& Lima (2013a,b), the effect of angular momentum in spherical collapse was also studied in dark energy cosmologies.}. 
%Apart 
We followed the standard recipe of SIMs, namely we studied the evolution of a spherical perturbation (divided into spherical shells) from the Hubble flow phase till it reached the maximum of expansion (turn-around). After the shell reaches the maximum of expansion, it collapses towards the center and will cross other shell (``shell-crossing"), and as a consequence energy will no longer be an integral of motion. In order to go on with the calculation, studying the shells dynamics, and to finally obtain the final density profile, we assumed that the central potential varies adiabatically, as is usually done (Gunn 1977; Filmore \& Goldreich 1984), such that we obtained the profile
%The collapse was studied by means of adiabatic invariants, after ``shell-crossing", to obtain the final density profile (Gunn 1977, Fillmore \& %Goldreich 1984)
\begin{equation}
\rho(x)=\frac{\rho_{ta}(x_m)}{f(x_i)^3} \left[1+\frac{d \ln f(x_i)}{d \ln g(x_i)} \right]^{-1}
\label{eq:dturnnn}
\end{equation}
being $f(x_i)=x/x_m(x_i)$ the collapse factor, $x_i$ the initial radius, and $x_m (x_i)$ the turn-around radius, 
given by
\begin{equation}
x_m=g(x_i)=x_i \frac{1+ \overline{\delta}_i}
{\overline{\delta}_i-(\Omega_i^{-1}-1)}
\end{equation}
where $\overline{\delta}_i$ indicates the average overdensity in the considered shell, and $\Omega_i$ the density parameter.
The baryonic fraction was obtained as in McGaugh et al. (2010),
\begin{equation}
f_d = (M_b/M_{500})/f_b=F_b/f_b
\end{equation}
being $f_b=0.17 \pm 0.01$ (Komatsu et al. 2009) the universal baryon fraction, $f_d$ the detected baryon fraction, namely the ratio among the known baryons to the amount of baryons that one expects from the cosmic baryon fraction,  
and $F_b=M_b/M_{500}$\footnote{$M_{500}$ denoted the mass in a radius $R_{500}$, within which the mean overdensity is 500 times larger than the critical density $\rho_c$. The virial mass $M_{vir}$ (see the following),  can be converted to $M_{500}$ using White (2001), Hu \& Kravtsov (2003) and Lukic
et al. (2009) method.} the baryonic fraction.

The random angular momentum, $j$, 
%arising from random velocities (Ryden \& Gunn 1987) 
is expressed in terms of the ratio
of the pericentric, $r_{min}$, to apocentric radii, $r_{max}$, $e=\left( \frac{r_{min}}{r_{max}} \right)$ (Avila-Reese et al. 1998), taking also into account the fact that eccentricity is a function of the dynamical state of the system 
\begin{equation}
e(r_{max}) \simeq 0.8 (r_{max}/r_{ta})^{0.1}
\end{equation}
for $r_{max}< 0.1 r_{ta}$, as shown by  Ascasibar, Yepes \& G\"ottleber (2004).

The ``ordered angular momentum", $h$, arising from tidal torques of large scale structure on the proto-structure, is obtained through the tidal torque theory (TTT) (Hoyle 1953; Peebles 1969; White 1984; Ryden 1988; Eisenstein \& Loeb 1995). 

The steepening effect produced by the DM compression caused by baryons collapse, the so-called adiabatic contraction (AC), was 
calculated according to the Gnedin et al. (2004) prescriptions. 
{As usual, it is assumed that the density profile of DM and baryons is the same (Mo et al. 1998; Keeton 2001; Treu \& Koopmans 2002; Cardone \&
Sereno 2005; IS11), and it is given by a NFW profile. The final distribution of baryons is assumed
to be a disk 
%(for spiral galaxies) 
(Blumenthal et al. 1986; Flores et al. 1993; Mo et al. 1998;
Klypin et al. 2002; Cardone \& Sereno 2005). For precision's sake, in our calculations, we shall assume Klypin et al. (2002) model for the baryons distribution (see their Section 2.1), {and their parameters: disc radial scale-length $r_{\rm d}=3-3.5$ kpc, and vertical scale-length $z_0=400$ pc. The bulge, having a Gaussian density $\rho \propto \exp (-s^2/2)$ is similar to a pseudo-bulge rather than a classical bulge with de Vaucouleurs profile (Kormendy \& Kennicut 2004). }
%, when dealing with mass scales typical of spiral galaxies. 
}.

Dynamical friction was taken into account as described in DP09. Its effect on structure formation is obtained by adding the dynamical friction force in the equation of motion (Eq. A14 of DP09).

The exchange of angular momentum between baryons and DM, through dynamical friction, becomes important in the later phases of structure formation, when baryons density increases due to the collapse, giving rise to a coupling process of the two components (Klypin et al. 2001; Klypin,  Zhao, and Somerville 2002).

Important physical processes like gas cooling, star formation, supernova feedback, and reionization were included as in Li et al. (2010) (Sect. 2.2.2, 2.2.3), and De Lucia \& Helmi (2008).
%, respectively.

Concerning reionization, as in Li et al. (2010), we used the Gnedin (2000) results, namely that reionization reduces the content of baryons in the case of haloes having a peculiar mass (``filtering mass" scale) depending on $z$. So, the baryon fraction changes as 
\begin{equation}
f_{b, halo} (z,M_{vir})= \frac{f_b}{[1+0.26 M_F (z)/M_{vir}]^3}
\end{equation}
where $M_{vir}$ is the virial mass, and $M_F$, the ``filtering mass", is given in Kravtsov et al. (2004). Reionization happens in the redshift range 15-11.5.

The cooling of the gas can be treated in different manners, following Ryden (1988), or as classical cooling flow (e.g., White \& Frenk 1991) (see Sect. 2.2.2 of Li et al. 2010).

Concerning star formation, the details are given in De Lucia \& Helmi (2008), while those of supernova feedback are described in Croton et al. (2006)
and De Lucia et al. (2004). 
Gas is distributed in an exponential disc, and star formation happens at a rate
\begin{equation}
\psi=0.03 M_{sf}/t_{dyn}
\end{equation}
where $M_{sf}$ is the mass of the gas above a critical density threshold, and $t_{dyn}$ the dynamical time of the disc, and the IMF is Chabrier one. 
If $\Delta t$ is the time-step 
the amount of new stars is
\begin{equation}
\Delta M_{\star}= \psi \Delta t
\end{equation}
 
Interstellar medium will receive a quantity of energy from supernovae, given by 
\begin{equation}
\Delta E_{SN}= 0.5 \epsilon_{halo} \Delta M_{\ast} V^2_{SN}
\end{equation}
where $\epsilon_{halo}=0.35$ (Li et al. 2010) is the reheating disc gas efficiency due to energy, and $V^2_{SN}=\eta_{SN} E_{SN}$ is the energy injection, per unit solar mass, by supernovae. $E_{SN}= 10^{51}$ erg, is the energy released by a supernovae explosion, on average, and $\eta_{SN}= 8 \times 10^{-3}/M_{\odot}$ is the supernovae number per solar mass obtained from a Chabrier initial mass function (IMF). 
The mass of cold gas reheated by supernovae can be assumed proportional to stars formed
\begin{equation}
\Delta M_{reheat} = \epsilon_{disc} \Delta M_{\ast}
\end{equation}
where $\epsilon_{disc}=3.5$ (Li et al. 2010). If this gas is re-added to the hot phase, this will
produce a change in the thermal energy by
\begin{equation}
\Delta E_{hot}= 0.5 \Delta M_{reheat} V^2_{vir}
\end{equation}
In the case $\Delta E_{SN} > \Delta E_{hot}$, the hot gas will be ejected from the halo, and can be assumed equal to 
\begin{equation}
\Delta M_{eject}= \frac{\Delta E_{SN}-\Delta E_{hot}}{0.5 V^2_{vir}}.
\end{equation}

The halo can accrete the ejected material that becomes part of the hot component related to the central galaxy (De Lucia et al.
2004; Croton et al. 2006).

In the present model, the perturbation is made of DM and baryons.
%% initially in the form of diffuse gas.????? 
The perturbation evolves till the maximum radius and then collapses, firstly in the DM component, and then in the baryon component. 
After a given shell reaches turn-around, it starts to collapse. Baryonic clumps exchange angular momentum with DM, cool forming stars, and accrete to the center, giving rise to the bulge (Immeli, Samland, Gerhard, \& Westera 2004; Lackner \& Ostriker 2010).

%
%The results of the previous model are in agreement with previous studies on the cusp flattening produced by heating of DM by collapsing clumps of baryons 
%(El-Zant et al. 2001, 2004; Romano-Diaz et al. 2008, and Cole et al. 2011), and by Supernovae feedback (Governato et al. 2010 (see DP11 Fig. 4); %Mashchenko et al. 2006, 2008).

\begin{figure}
\psfig{file=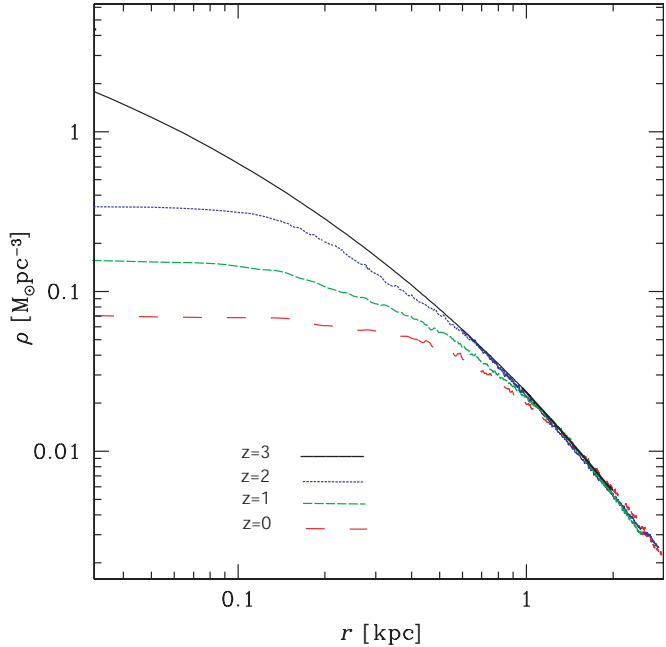,width=10.0cm}
%\psfig{file=saaa.eps,width=8.0cm}
%\psfig{file=deeprof.ps,width=8.0cm}
%\psfig{file=deeprof.ps,width=8.0cm}
%\psfig{file=dprofok.ps,width=8.0cm}
%\psfig{file=dprofok4.ps,width=8.0cm}
%\psfig{file=deeprof.ps,width=8.0cm}
%\picplace {2.0cm}
\caption[]{Plot of the density profile of a DM halo having a virial mass $M_{vir}=10^{9} M_{\odot}$. The evolution at $z=3$, $z=2$, $z=1$, $z=0$, is represented by the solid line, dotted line, short-dashed line, and long dashed-line, respectively. }
\end{figure}

\begin{figure}
\centerline{\hbox{(a)
\psfig{file=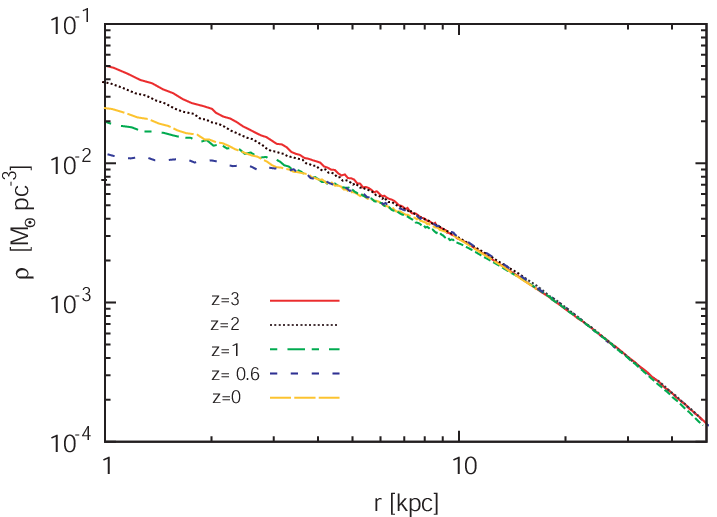,width=10.0cm}
}}
\centerline{\hbox{(b)
\psfig{file=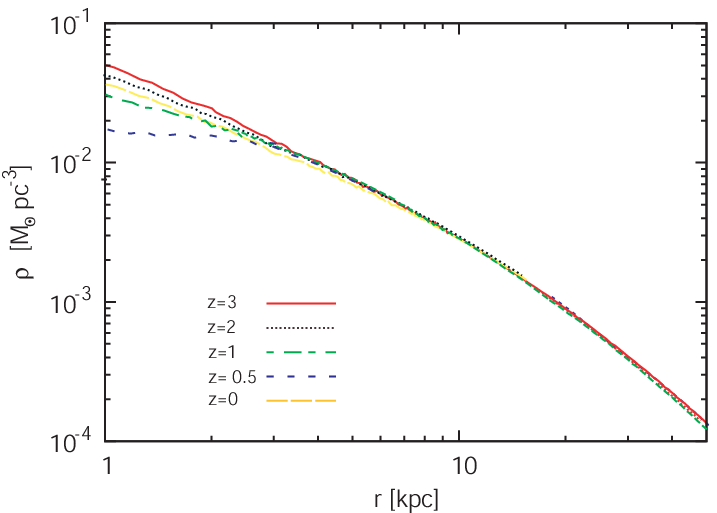,width=10.0cm}
}}
%\psfig{file=saaa.eps,width=8.0cm}
%\psfig{file=deeprof.ps,width=8.0cm}
%\psfig{file=deeprof.ps,width=8.0cm}
%\psfig{file=dprofok.ps,width=8.0cm}
%\psfig{file=dprofok4.ps,width=8.0cm}
%\psfig{file=deeprof.ps,width=8.0cm}
%\picplace {2.0cm}
\caption[]{Top panel: plot of the density profile of a galaxy having a virial mass $M_{vir}= 10^{11} M_{\odot}$
%. In the top panel, the 
and bulge mass of $4.5 \times 10^{9} M_{\odot}$. The evolution at $z=3$, $z=2$, $z=1$, $z=0.6$, $z=0$, is represented by the solid line, dotted line, long-short-dashed line, dashed line, and long dashed-line, respectively. In the bottom panel, we plot the evolution of a DM halo having a virial mass $M_{vir}=10^{12} M_{\odot}$, and same bulge mass as the previous halo. In this case, the dashed line represents the evolution of the profile at $z=0.5$.}
\end{figure}

\begin{figure}
\centerline{\hbox{(a)
\hspace{-0.5cm}
\psfig{file=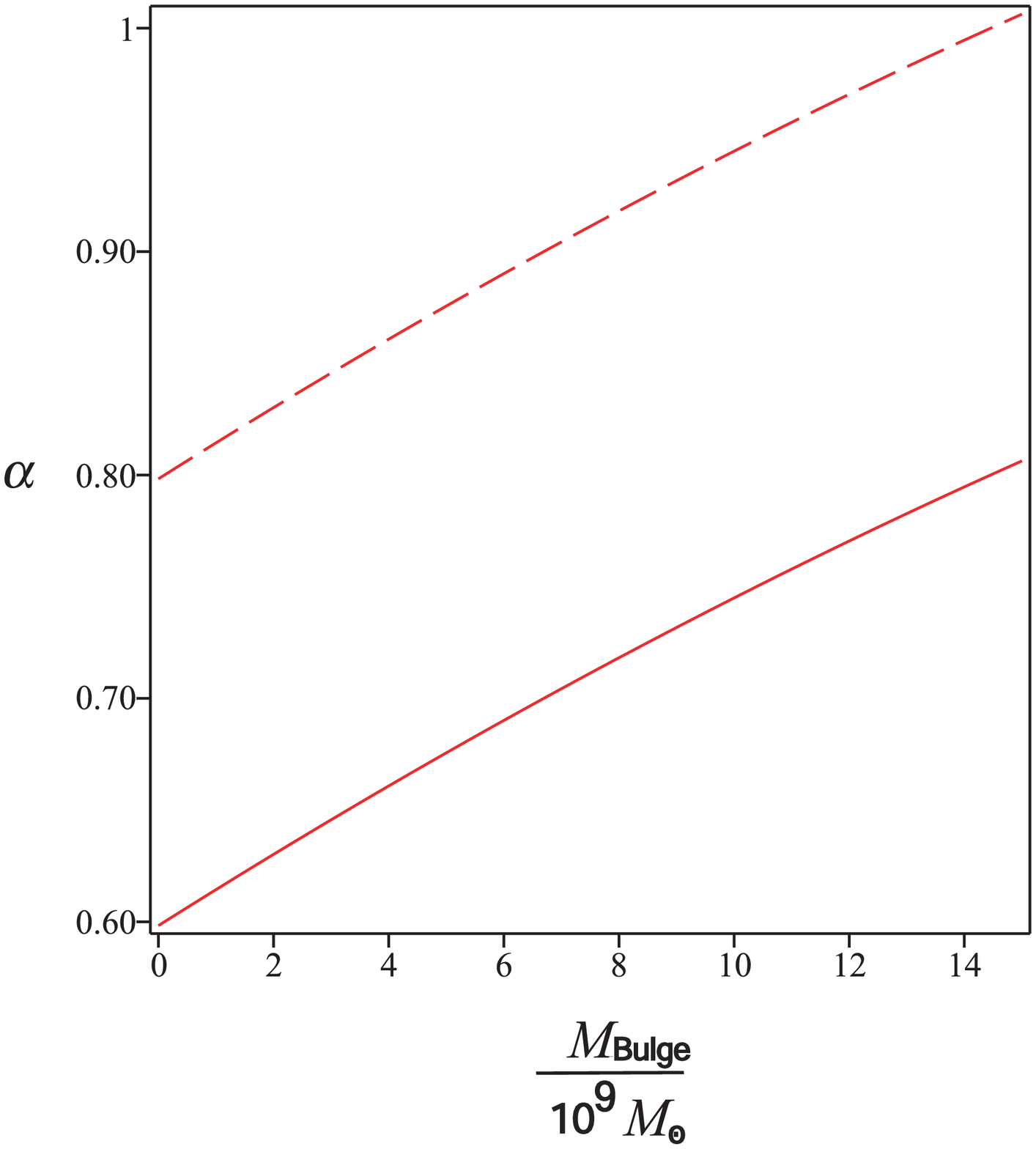,width=9.0cm}
}}
\centerline{\hbox{(b)
\psfig{file=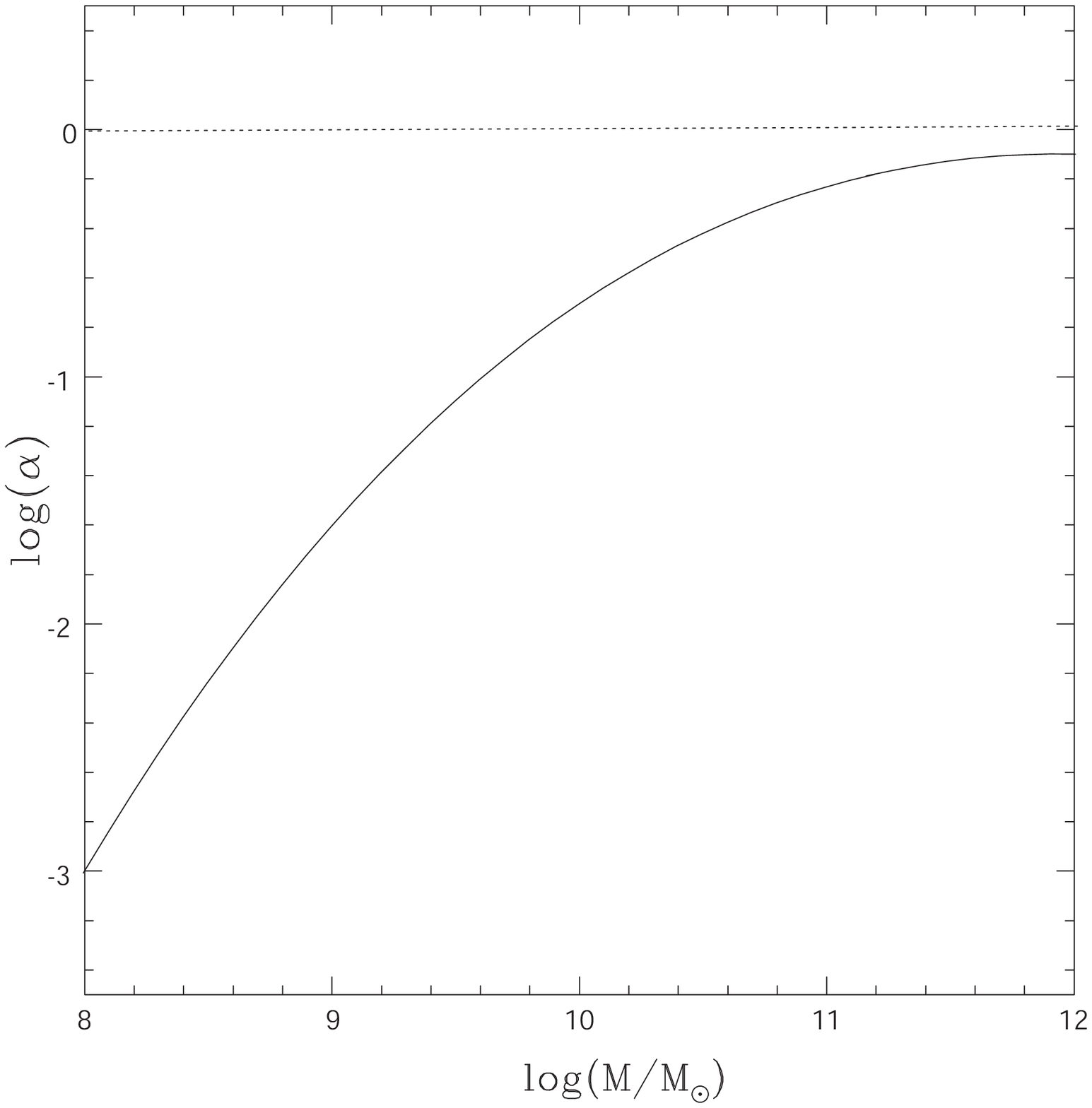,width=9.0cm}
}}
%\psfig{file=saaa.eps,width=8.0cm}
%\psfig{file=deeprof.ps,width=8.0cm}
%\psfig{file=deeprof.ps,width=8.0cm}
%\psfig{file=dprofok.ps,width=8.0cm}
%\psfig{file=dprofok4.ps,width=8.0cm}
%\psfig{file=deeprof.ps,width=8.0cm}
%\picplace {2.0cm}
\caption[]{Top panel: plot of the logarithm of the slope of $\alpha$ in terms of the bulge mass for two different virial masses: a)
virial mass equal to {$10^{11} M_{\odot}$ (solid line), and b) virial mass equal to $10^{12} M_{\odot}$ (dashed line)}.
%. The solid line correspond to the case the galaxy has no bulge, the short, and dashed lines correspond to the case the bulge has %a mass $4.5 \times 10^{9} M_{\odot}$, and $6.5 \times 10^{9} M_{\odot}$, respectively.
Bottom panel: plot of the logarithm of the slope $\alpha$ in terms of the virial mass.
%for three different bulge masses: a) no bulge mass (solid line); b) bulge mass equal to $4.5 \times 10^{9} M_{\odot}$ (short %dashed line), and c) bulge mass equal to $6.5 \times 10^{9} M_{\odot}$ (long dashed line).
}
\end{figure}

%%\begin{figure}
%%\centerline{\hbox{(a)
%%\psfig{file=sa05.eps,width=4.0cm} (b)
%%\psfig{file=sa4.eps,width=4.0cm}
%%}}
%%\centerline{\hbox{(c)
%%\psfig{file=sa5.eps,width=4.0cm}
%%\psfig{file=saaaa.eps,width=4.0cm}
%%}}
%\psfig{file=saaa.eps,width=8.0cm}
%\psfig{file=deeprof.ps,width=8.0cm}
%\psfig{file=deeprof.ps,width=8.0cm}
%\psfig{file=dprofok.ps,width=8.0cm}
%\psfig{file=dprofok4.ps,width=8.0cm}
%\psfig{file=deeprof.ps,width=8.0cm}
%\picplace {2.0cm}
%%\caption[]{}
%%\end{figure}

\section{Results and discussion}

The results of the calculation are shown in Fig. 1-3. In Fig. 1, we plot the density profile evolution of a bulgeless dwarf galaxy with virial mass $M_{vir}=10^{9} M_{\odot}$, at different $z$. Baryons are initially distributed as DM (see Lackner \& Ostriker 2010) with a baryon fraction $f_d=0.04$, and characterized by a spin parameter $\lambda=0.05$ (see DP12b).\footnote{The spin parameter is defined as
\begin{equation}
\lambda= L \sqrt{|E|}/GM^{5/2}
\end{equation}
(Peebles 1969), where E indicates the system internal energy, $L$ the angular momentum, and $M$ the structure mass. 
The spin parameter $\lambda$ has a log-normal distribution having a maximum at $\lambda = 0.035$ and a 90\% probability 
to find it in the 0.02-0.1 range (Vivitska et al. 2002).
}

%UGC 6446, having h ' 400 kpc km /s ( ' 0.05), and the baryon fraction (see Sec. 2) is fd ' 0.04.
The galaxy evolution was started at $z=50$, and Fig. 1 plots the dwarf density profile at $z=3$ (solid line), $z=2$ (dotted line), 
$z=1$ (short-dashed line), and $z=0$ (long-dashed line). After the system virializes around $z=10$, in the initial evolution stage adiabatic contraction (AC) at $z \simeq 5$ produces a steepening of the profile (not shown in the figure).

The dwarf is bulgeless in agreement with simulations (e.g., Governato et al. 2010) and observations, showing that usually dwarfs have no bulge. 

The flattening of the profile observed is due to the interplay between DM and baryons, secondary infall, and two-body relaxation. In the infall, baryons clumps interact with DM particles, heating them with the result that they move on external orbits decreasing the inner density. A core forms at the center of the dwarf starting from $z \simeq 3$ and enlarges till $z=0$, when the core radius is $r \simeq 1$ kpc. No cusp is observed to reform in the late stage of evolution, like in {IS11, but this is expected since in this case
%in our case 
there is no bulge, differently from their simulation}.  
The previous result, as already discussed in DP09,  DP12a,b, is in agreement with the Romano-Diaz et al. (2008) simulations. Similarly to other studies (e.g., El-Zant et al. 2001, 2004; Cole et al. 2011) the profile flattening was interpreted as being due to clumps falling to the galaxy center, and DM heating by dynamical friction. In DP12a, we carried a comparison with the Governato et al. (2012) results concerning the formation of a bulgeless dwarf, finding a good agreement.  

{In Fig. 2a, 2b, we study the effect of a bulge on galaxies of different mass. In order to compare the result with IS11 simulations, a bulge is added to the {cored} DM halo.

In the quoted figures, we plot the evolution of a density profile of a galaxy with $M_{vir} =10^{11} M_{\odot}$ (panel a), and 
of $M_{vir} =10^{12} M_{\odot}$ (panel b), both having a bulge mass of $4.5 \times 10^{9} M_{\odot}$. }
%, and $6.5 \times 10^{9} M_{\odot}$, respectively. 
In this case the baryon fraction is $f_d=0.17$ ($F_b=0.03$) (see McGaugh et al. 2010; Dai et al. 2012) and $\lambda = 0.05$ (note that in IS11, $\lambda = 0.1$).  
%{\bf PIU' PICCOLO.}
%Tonini, C.; Lapi, A.; Shankar, F.; Salucci, P., 2006, ApJ 638,..13	

%Baryons are initially distributed as DM (see Lackner \& Ostriker 2010) with a baryon fraction $f=$, and characterized by %$\lambda???$. 
{The density profile is plotted at $z=3$ (solid line), $z=2$ (dotted line), $z=1$ (long-short-dashed line), $z=0.6$ (dashed line) (in the case of $M_{vir} =10^{11} M_{\odot}$ halo), $z=0.5$ (in the case of $M_{vir} =10^{12} M_{\odot}$ halo), and $z=0$ (long-dashed line). 
}
%
%%As for the dwarf, this galaxy is subject to AC in its early stage, and then the profile starts to flatten.
%
%, similarly to the dwarf galaxy case. 

{In the case of the $M_{vir} =10^{11} M_{\odot}$ halo (panel a), the core enlarges till a redshift $z=0.6$, when the slope $\alpha$ reaches its minimum value, while in the case of the $M_{vir} =10^{12} M_{\odot}$ halo (panel b) $\alpha_{min}$ is reached at $z=0.5$. After the quoted redshifts, the profile starts to steepen and the inner slope reaches a value of $\alpha \simeq 0.65$ in the case of the $M_{vir} =10^{11} M_{\odot}$ DM halo, and $\alpha \simeq 0.85$ in the case of the $M_{vir} =10^{12} M_{\odot}$ DM halo. In the case in which no bulge is present the values of the inner slope are $\alpha \simeq 0.6$, and $\alpha \simeq 0.8$, respectively.

The flattening process giving rise to the formation of a core, and the following steepening  happens in a similar way to the IS11 simulation. 
%After $z=1$, the profile steepens in a similar way to what shown in Inoue \& Satoh (2011), but the steepening is much less than %what was showed by the previous authors. 
The result shows that the bulge formation has an important role in the density profile settling. 
%slightly steepens the profile. 
This is due to the deepening of the potential well as bulge forms, dragging matter to the center. 
%Fig. 2b, shows the density profile evolution for a galaxy having a mass of $10^{11} M_{\odot}$ as the previous one, as reported, for a larger bulge %mass.  
%
%but with a larger bulge ($4.5 \times 10^{9} M_{\odot}$ 
%The inner DM slope of the galaxy in fig. 2a has a slope $\alpha \simeq 0.65$, and $\alpha \simeq 0.55$ in absence of the bulge.
%In Fig. 2b, we plot the case of a $M_{vir}= 10^{12} M_{\odot}$ galaxy, having the same bulge mass ($4.5 \times 10^{9} M_{\odot}$) %of the previous halo. The evolution is similar to what described in the previous Fig. 2a, but the final density profile is %steeper. In this case, the slope is $\alpha \simeq 0.85$, and $\alpha \simeq 0.75$ in bulge absence.

Fig. 2a,b, shows two important issues: a) in the presence of a bulge, the final density profile is steeper than in the case it is absent. As we show in the following of the paper, there is a correlation among the bulge mass and the inner slope of the density profile; b) galaxies having the same bulge mass but larger virial mass have steeper inner slopes (as already discussed in DP09). 
%%more massive galaxies have steeper inner slopes (as already discussed in DP09). 
%As observed, in Fig. 2b, 
%the final profile is steeper than the previous case (Fig. 2a), showing that the larger is the bulge formed the steeper the %density profile. At the same time, the $10^{11} M_{\odot}$ galaxy, has a steeper inner profile than the dwarf galaxy, %independently from the bulge formation (see the following). 
%In this case, the inner DM slope is $\alpha \simeq 0.7$.
}

The reason of the steepening of the inner density profile with mass, 
%always stressed comparing the dwarf galaxy profile with that of the $10^{11} M_{\odot}$ galaxy, 
was already explained in DP09. In a few words, less massive proto-structures 
are born from smaller peaks of the density field (Del Popolo \& Gambera 1996). Since angular momentum is anti-correlated with the peak height (Hoffman 1986), the smaller an object is the larger the angular momentum it acquires. As a consequence, particles in the proto-structure are endowed with larger angular momentum and will stay for longer times far away from the proto-structure center, producing a shallower profile (see also William et al. 2004). In DP12a,b, it was shown, that a
correlation among the density profile shape, the baryon content of the structure, and the angular momentum acquired, exists. 
%{\bf As described in DP12a,b, exists a correlation between the angular momentum acquired by the structure, the baryon content of the structure, and the %shape of the density profile. }

{In order to understand better if exists a correlation among bulge mass and inner slope of the density profile, in Fig. 3, we studied how the inner slope changes when the bulge mass and virial mass is changed. 
Fig. 3a plots the inner slope $\alpha$ in terms of the bulge mass for a DM halo having a mass $10^{11} M_{\odot}$ (solid line), and in for a DM halo having a mass $10^{12} M_{\odot}$ (dashed line). According to Weinzirl et al. (2009), a large fraction
($\simeq 69\%$) of bright spirals have a ratio among the bulge mass and total mass $\leq 0.2$. So, we studied the slope change in the bulge mass range $M_{Bulge}=0-10 \times 10^9 M_{\odot}$, for a halo of $10^{11} M_{\odot}$, and a halo of $10^{12} M_{\odot}$. 

The plot shows an increase of the slope with the bulge mass, and virial mass. In Fig. 3b, I plot the inner slope in terms of the virial mass.
%, for different bulge mass, namely a) no bulge mass; b) a bulge mass of $4.5 \times 10^{9} M_{\odot}$ (short dashed line), and c) %a bulge mass of $6.5 \times 10^{9} M_{\odot}$ (dashed line). 
The plot shows again an increase of the inner slope with virial mass. 
%and bulge mass. 
The result confirms previous results (Del Popolo 2010; Del Popolo 2011) on the universality of the density profiles. Real systems, constituted by DM and baryons have not universal density profiles. 

The previous results are also important in connection with the indirect detection of DM\footnote{Indirect detection tries to detect DM through its annihilation products (e.g., gamma rays, electrons, positrons, antiprotons, and neutrinos). }
in the galactic center and in the center of other galaxies. The annihilation flux is proportional to the square of the DM density, $\rho_{DM}^2$.

The degree of ``cuspiness" of a density profile is of fundamental importance in the probability of detecting the quoted annihilation flux from DM\footnote{The averaged annihilation flux on the solid angle $\Delta \Omega$, for a Moore's profile, having $\rho \propto r^{-1.5}$ is a factor $10^3$ larger than the flux from a NFW profile.} (see Hooper 2009, Del Popolo 2007; Del Popolo 2013).

}

We want to recall that the quoted results of cored profiles in dwarfs is in agreement with observational results (Burkert 1995; de Blok, Bosma, \& McGauch 2003; Swaters et al. 2003; Oh et al. 2010, 2011; Kuzio de Naray \& Kaufmann 2011), and simulations (e.g., Mashchenko et al. 2006, 2008; Governato et al. 2010). Also the steepening of the profile for objects of larger mass is in agreement with the density profiles of THINGS galaxies (de Blok et al. 2008). As already reported in the Introduction, the density profile of galaxies with brightness $M_B<-19$ are equally well described by cored or cuspy profiles, while this is not true for fainter galaxies ($M_B>-19$), which are better described by cored profiles. 
{Finally, the steepening of the density profile with increasing bulge mass is confirming Inoue \& Saitoh (2011) results, showing that bulge formation gives rise to a steepening of the profile. 
}

\section{Discussion and Summary}

In the present paper, we discussed the cusp-core problem, with particular attention to the role of the bulge formation on the final density profile. 
We have seen, that dwarf galaxies are bulgeless and as shown in Fig. 1, 
%that a bulgeless dwarf galaxy is 
that they are characterized by a final cored density profile, well fitted by a Burkert profile, in agreement with several previous studies based on a similar mechanism to this paper (El-Zant et al. 2001, 2004; Romano-Diaz et al. 2008; Goerdt et al. 2010; Cole et al. 2011) or a different one (e.g., Governato et al. 2010). 
In Fig. 2-3, we showed that:
%two different things: 
a) bulge formation influence the density profiles,
%in the presence of a bulge, the density profile steepens is slightly influenced by the bulge formation, 
with final profiles steeper than the profiles in absence of the bulge. b) Lower mass galaxies are characterized by flatter profiles than larger mass ones. This last point was also shown in DP09, and we summarized briefly why this is the case.
 
The inner slopes are steeper in the presence of a bulge, and the slope increases with the bulge mass, in agreement with IS11. At the same time more massive galaxies have steeper profiles, so that larger is the virial mass and bulge mass of a galaxy the steeper is its inner density profile.
%but at the same time more massive galaxies have steeper profiles, and the role of mass increase is larger than that of the increase of the bulge mass. 
We know that the total mass is almost constant, or slightly increasing between SO and Sc galaxies (Kormendy 1982), and that the bulge mass is decreasing (Weinzirl et al. 2009) from SO to Sc galaxies. So combining the two effects, previously discussed, an early type galaxy should have a steeper density profile with respect to late ones, and late type a flatter profile. So, for example one should observe different slopes even in the two different kinds of known Low Surface Brightness (LSB), galaxies  \footnote{LSB galaxies come into two different types: a) late type, gas-rich, DM dominated, dominated by an exponential disc, and bulgeless. 2) Early-type, bulge-dominated (Pickering et al. 1997).}, the late type ones, bulgeless, should have flat density profiles (as observed), and the early-type ones, bulge-dominated, should show a steeper profile. 
{A comparison of the density profile of these two types of LSBs would be very helpful in understanding better the cusp-core problem.
}

{Comparing our result with that of IS11, we may tell that our result is in agreement with their. 
%from the qualitative point of view there is an agreement, but from the quantitative one there is not. 
In our calculation, the effect of bulge in the density profile is clearly that steepening it. Our results, show a clear, strong correlation among the inner slope and the bulge mass.  
%formation influence the slope of the inner density profile, but the slope change is much slighter than the Inoue \& Satoh (2011) %paper. So our result is similar to that of Fu et al. (2007), namely that bulge formation has a small role on DM halo.
}

\section*{Acknowledgements}

We would like to thank the International Institute of Physics in Natal for the facilities and hospitality, and Charles Downing from Exeter University for a critical reading of the paper. 

{}

\end{document}